\documentstyle[aps,prb,twocolumn,psfig]{revtex} 
\begin{document} 
\draft 
\twocolumn[\hsize\textwidth\columnwidth\hsize\csname
@twocolumnfalse\endcsname
\title{ Momentum distribution of liquid helium } 
\author{Saverio  Moroni$^{(1)}$, Gaetano Senatore$^{(2,3)}$ and Stefano
 Fantoni$^{(1,4)}$}
 
\address{$^{(1)}$ International Centre for Theoretical Physics, I-34014, 
Trieste, Italy\\ 
$^{(2)}$Istituto Nazionale di Fisica della Materia\\ 
$^{(3)}$Dipartimento di Fisica Teorica, Universit\`a di Trieste, 
Strada Costiera 11,  I-34014 Trieste, Italy\\ 
$^{(4)}$Interdisciplinary Laboratory, SISSA, Via Beirut 2/4, I-34014, 
Trieste, Italy 
}
\date{Submitted to Phys. Rev. B,  July 18, 1996}
\maketitle 
\begin{abstract} 
 We have obtained the
one--body density matrix and the momentum distribution $n(p)$ of
liquid $^4$He at $T=0^o$K from Diffusion Monte Carlo (DMC)
simulations, using trial functions optimized via the Euler Monte Carlo
(EMC) method. We find a condensate fraction smaller than in previous
calculations. Though we do not explicitly include long--range
correlations in our calculations, we get a momentum distribution at long
wavelength which is compatible with the presence of long--range
correlations in the exact wave function. We have also studied $^3$He,
using fixed--node DMC, with nodes and trial functions provided by the
EMC. In particular, we analyze the momentum distribution $n(p)$ with
respect to the discontinuity $Z$ as well as the singular behavior, at
the Fermi surface.  We also show that an approximate factorization of
the one-body density matrix $\rho(r)\simeq \rho_0(r)\rho_B(r)$ holds,
with $\rho_0(r)$ and $\rho_B(r)$ respectively the density matrix of
the ideal Fermi gas and the density matrix of a Bose $^3$He. 
\end{abstract}
\pacs{PACS numbers: 67., 67.40.-w, 67.55.-s}
\vskip2pc]

\renewcommand{\thepage}{\hskip 8.9cm \arabic{page} \hfill Typeset
using REV\TeX }

\narrowtext

\section{Introduction}

The momentum distribution $n(p)=<a^{\dagger}_{\bf p}a_{\bf p}>$ is a
fundamental quantity for the study of both the static and the
dynamical properties of quantum liquids, as it gives direct
information on the high momentum components of the ground state wave
function\cite{momdis89}. Experimentally, access to $n(p)$ is provided
by deep inelastic neutron scattering at large momentum transfer $\hbar
Q$. The extraction of $n(p)$ from the measured scattered intensity,
however, is affected by the the limitations imposed by the
experimental resolution and the final-state interactions. Thus, the
most accurate information on the momentum distribution of $^4$He
available to date is likely to be the one obtained through accurate,
microscopic calculations, such as those presented in this study.

At $T=0^o$K the momentum distribution $n(p)$ of an ideal Bose gas is
given by a delta function $\delta({\bf p})$, corresponding to all particle
being in the condensate.  On the contrary, the condensate fraction of
$^4$He at the equilibrium density is less than $10\%$, implying that
the effects of the strongly repulsive core of the inter-atomic
interaction is non perturbative.  Similarly for $^3$He the
discontinuity of $n(p)$ at the Fermi momentum $\hbar p_F$, which gives
the strength of the quasiparticle pole, is $\sim 0.2$ at equilibrium
instead of $1$, as in the ideal Fermi gas. In fact, it has been always
difficult to carry out {\it ab initio} calculations of the momentum
distribution for these systems, within the field--theoretical
approach\cite{agd63}.

Modern, realistic, quantitative calculations only started with the
development of Variational Monte Carlo (VMC) methods\cite{Ceperley79}
and the Hyper--Netted--Chain (HNC,FHNC)
equations\cite{Fantoni78,Ristig81} for Jastrow models of both Bose and
Fermi liquids.  FHNC and HNC equations for the momentum distribution
have been numerically solved in Variational calculations of nuclear
matter\cite{Fantoni84}, liquid $^4$He\cite{Manousakis85,momdis89}, and
liquid $^3$He\cite{Fabrocini82,Flynn86,Fabrocini92}, for which triplet
and backflow correlations were also taken into account.  To improve
upon the above variational estimates of $n(p)$, non conventional
perturbative techniques, based on correlated basis
functions\cite{Feenberg69} (CBF), have been developed, and applied to
liquid $^4$He\cite{Manousakis85} and nuclear
matter\cite{Fantoni84,momdis89}.

HNC and FHNC theories have the merit to allow for fine details of
interparticle correlations, such as  long--range behavior,  spin
dependence, anisotropies in inhomogeneous systems. However, not
all cluster diagrams resulting from the theory and involved in the HNC
formalism, can be summed in closed form and in FHNC the procedure to
estimate the elementary diagrams with exchange bonds is not completely
under control.  Therefore, approximations like scaling\cite{Usmani82},
interpolation\cite{Fabrocini82} or truncated
summations\cite{Smith76,Krotscheck86,Wang93} must be invoked, and, in
liquid helium calculations, these result in a non negligible loss of
accuracy with respect to a fully variational
treatment\cite{Moroni95a,Moroni95b}.

The momentum distribution of liquid $^4$He and $^3$He has also been
calculated, at zero temperature, by using the Green Function Monte
Carlo (GFMC) method\cite{Whitlock87,momdis89}. At finite temperature
calculations have been performed for $^4$He by Path Integral Monte
Carlo (PIMC) \cite{Ceperley86,momdis89} and, more recently, by VMC
with trial functions of the Shadow type\cite{Masserini92}.  The
available theoretical estimates of the momentum distribution of liquid
$^4$He provided by variational and GFMC methods are in reasonably good
agreement among themselves, except for low momenta and for the
condensate fraction. For liquid $^3$He the situation is less
satisfactory.

The GFMC and the Diffusion Monte Carlo (DMC)
methods\cite{Reynolds82,Ceperley91,Moroni95b}, afford to date the most
precise tools to perform ground--state calculations for many--particle
systems. For Bosons they provide estimates of the energy that are
virtually exact, within the statistical accuracy.  In fact the same is
true for the averages of operators that are (R--)diagonal (not the
case for $n(p)$!), for which algorithms exist, like the so--called
forward walking method\cite{lkc74,cb95}, that yield {\it pure}
estimates. On the other hand, the estimates of observables that do not
commute with the Hamiltonian are usually obtained from {\it mixed}
averages---through an extrapolation procedure whose accuracy depend on
the quality of the trial function $\Psi$ used for the importance
sampling\cite{Ceperley79}. The extrapolation introduces a bias in the
estimates which is second order in the difference between $\Psi$ and
ground--state wave function $\Phi_0$.
For Fermions there is an additional
source of error related to the existence of the so--called {\it sign
problem}. To date, to get a numerically stable algorithm it is
customary to approximate the unknown nodes of the sought ground state
$\Phi_0$ with those of the trial function $\Psi$ ( {\it fixed--node}
approximation\cite{Ceperley91}). This imposes a bias on any  average, which
for non-diagonal observables cumulates with the one arising from the
extrapolation procedure. Therefore, to minimize systematic errors,
especially in the evaluation of properties such as the momentum
distribution, it is necessary to achieve maximum accuracy in the
optimization of the trial function.

Recently, a new optimization procedure based on Monte Carlo
calculations and denoted as Euler Monte Carlo (EMC) method has been
proposed\cite{Vitiello92,Moroni95b}. This EMC method has been
successfully applied to both liquid $^4$He and
$^3$He\cite{Moroni95a,Moroni95b}. The EMC wave functions have pair and
triplet correlations fully optimized and provide the lowest available
energy upperbounds. Moreover, their use in Diffusion Monte Carlo (DMC)
calculations\cite{Reynolds82,Ceperley91,Moroni95b} has led to results
of unprecedented accuracy for the energy, pair function and static
structure function.

In this paper we present results for the one--body density matrix
$\rho(r)$ and the momentum distribution $n(p)$ of liquid $^4$He and
$^3$He, at various densities, obtained with DMC calculations based on
EMC wave functions.  The plan of the paper is as follows. In the
next Section we summarize the computational details involved in
the calculation of $n(p)$. We then present the results for liquid
$^4$H in Sec. III and those for liquid $^3$He in Sec. IV. We finally
offer a summary and conclusions in Sec. V.

\renewcommand{\thepage}{\arabic{page}}

 \section{Computational details}

In DMC simulations\cite{Reynolds82,Ceperley91} the ground state wave
functions $\Phi_0$ is sampled through a random walk in configuration
space, guided by a trial function $\Psi$. In practice one samples the
mixed probability $f=\Phi_0\Psi$.  We chose $\Psi=SF$ for
$^4$He and $\Psi=D_{\uparrow}D_{\downarrow}F$ for $^3$He, with the
{\it correlation} part $F$ symmetric in the particle coordinates, $S$
a symmetrized product of one-particle orbitals and $D_{\uparrow}$ and
$D_{\downarrow}$ Slater determinants of one-particle orbitals for
particles of up and down spin. In the homogeneous liquid $S=const$ and
$D_{\sigma}$ is built from plane waves (PW), or from plane waves with
 short (SBF) or long (LBF) ranged back-flow corrections.  As a full
account has already been given elsewhere\cite{Moroni95a,Moroni95b} of
both the EMC method, which we employ to construct and optimize the
trial wave function $\Psi$, and of the use of EMC wave functions in
DMC simulations, here we shall restrict to essential details.

All the calculations (variational and diffusion) presented in this
paper, unless explicitly noted, have been performed using EMC
wave functions with fully optimized pair and triplet
pseudopotentials\cite{Moroni95b} (OJOT)and modeling Helium with the HFDHE2
pair potential of Aziz {\it et al.}\cite{Aziz79}.  A cubic
simulation box and periodic boundary conditions were used.  For $^3$He
backflow correlations have  been included\cite{Moroni95b}, in the
usual way\cite{Schmidt81,Manousakis83,Kwon93}, by replacing the plane
waves $\exp\left(i{\bf k}_{i}\cdot{\bf r}_j\right)$ in the Slater
determinants with $\exp\left(i{\bf k}_{i}\cdot{\bf x}_j\right)$, where
${\bf x}_j={\bf r}_j+\sum_{k\neq j} \eta(r_{jk})({\bf r}_{j} -{\bf
r}_k)$.  The function $\eta(r)$ is taken either short--ranged
\cite{Schmidt81} (SR)
\begin{equation} 
\eta_S(r) =\lambda_B\exp\left(-(r-r_B)^2/\omega_B^2\right)((2r-L)/L)^3, 
\label{backflow}
\end{equation}
or  long--ranged\cite{Panoff89} (LR)
\begin{equation}
\eta_L(r) = \lambda_B\exp\left(-(r-r_B)^2/\omega_B^2\right)+
\lambda^{\prime}_B/r^3,
\label{backflowL}
\end{equation}
with $\lambda_B$, $r_B$, $\omega_B$ and $\lambda^{\prime}_B$
variational parameters. The long--ranged backflow function is smoothly
cutoff at the the boundary of the simulation box of size $L$ by
replacing the expression given in Eq. (\ref{backflowL}) with
$\eta_L^{\prime}(r)=\eta_L(r)+\eta_L(L-r)-2\eta_L(L/2)$. In practice,
we first simultaneously optimize the pair and triplet pseudopotentials
with the backflow parameters as specified in
Refs. \onlinecite{Schmidt81} and \onlinecite{Panoff89}, and then we
optimize $\eta(r)$ at fixed pseudopotentials. 

The $^3$He results presented below  were obtained with
 short--ranged backflow, unless otherwise specified. Also the DMC
simulations were performed within the fixed--node approximation,
whereby the nodes of the ground--state $\Phi_0$ are assumed to
coincide with those of the EMC trial function.

In a uniform liquid  in a state described  by the wave function  $\Psi$ the
one--body density matrix  can be
defined as
\begin{eqnarray}
&&\rho(r)= \cr \cr
&&\frac{V\int\ d{\bf r}_2\cdots d{\bf r}_N
\Psi^{\star}({\bf r}_{1},{\bf r}_2,\cdots,{\bf r}_N)
\Psi({\bf r}_1+{\bf r},{\bf r}_2,\cdots,{\bf r}_N)}
 {\int\ d{\bf r}_1\cdots d{\bf r}_N|\Psi({\bf r}_1,{\bf
r}_2,\cdots,{\bf r}_N)|^2}, \cr
&&    
\label{dmatrix1}
\end{eqnarray}
so that, having imposed periodic boundary conditions to the $N$
particles in the volume $V$,
\begin{equation}
\rho(0)=1,
\label{norm}
\end{equation}
and the independence on ${\bf r}_1$ follows from
translational invariance. For Fermions, the integration over the
variable ${\bf r}_i$ is understood to imply also the trace over spin
projection. If we denote with $R$ and $R'$ respectively the
configurations $({\bf r}_1,{\bf r}_2,\cdots,{\bf r}_N)$ and $({\bf
r}_1+{\bf r},{\bf r}_2,\cdots,{\bf r}_N)$, and we exploit the
translational invariance, we can rewrite Eq. (\ref{dmatrix1}) as
\begin{eqnarray}
\rho(r)=\frac{\int d R\Psi^{\star}( R)
\Psi( R')}
 {\int d R|\Psi( R)|^2}
=\int d RP(R) \frac{\Psi(R')}{\Psi(R)},
\label{dmatrix2}
\end{eqnarray}
with 
\begin{eqnarray}
P(R)=\frac{|\Psi(R)|^2} {\int d R|\Psi( R)|^2}
\label{prob}
\end{eqnarray}
the probability induced by the wave function $\Psi$. Using
Eq. (\ref{dmatrix2}), the variational density matrix (i.e., the one
defined in term of the trial wave function $\Psi$)  may be conveniently calculated by
Monte Carlo as
\begin{eqnarray}
\rho_{\rm VMC}(r)\simeq \frac{1}{\cal N} \sum_{i=1}^{\cal N} \frac{\Psi(R'_i)}{\Psi(R_i)},
\label{rhoT}
\end{eqnarray}
 with the configurations $R_i$ drawn from the probability $P(R)$.   
The DMC extrapolated estimate  $\rho(r)$ is given by
\begin{eqnarray}
\rho(r)=2\rho_{\rm mix}(r)-\rho_{\rm VMC}(r),
\end{eqnarray}
where the mixed estimate  $\rho_{\rm mix}(r)$ is
calculated from an expression identical with that of Eq. (\ref{rhoT}),
with the configurations $R_i$ drawn however from the mixed probability
$P_{\rm mix}(R)=f(R) /\int d R f(R)$, $f(R)=\Phi_0(R)\Psi(R)$.
In practice, the auxiliary configurations $R'$ appearing in
Eq. (\ref{rhoT}) are generated from a given $R$ by moving a particle
either of fixed increments along a random direction (FM) or to points
randomly distributed in the simulation box (RM). It turns out that the
two methods give more accurate results at small and at large $r$,
respectively.

The momentum distribution is defined as $n(p)=\langle a^{\dagger}_{\bf
p}a_{\bf p}\rangle$, where an average on spin projections is also
implied for unpolarized $^3$He.  It is simply related to the one--body
density matrix, by a Fourier transform:
\begin{eqnarray}
n(p) &=&\frac{\rho}{\nu}\int d{\bf r}
  e^{i{\bf p}\cdot{\bf r}}\rho(r)  \cr
 &=&  \frac{\rho}{\nu}\left(n_0\delta({\bf p})+ \int d{\bf r} e^{i{\bf p}\cdot{\bf r}}\left(\rho(r)-n_0\right)\right),
 \label{momentum}
\end{eqnarray}
with $\rho=N/V$ the density of the system, $\nu$ the degeneracy
factor, which is $1$ for $^4$He and the fully polarized $^3$He, and
$2$ for normal $^3$He, and $n_0$  the large $r$ limit of the density
matrix, $n_0=\rho(\infty)$. In fact $\rho(r)$ vanishes for large
values of $r$ in $^3$He, whereas for $^4$He it saturates to $n_0\neq
0$, due to macroscopic occupation of the state with zero
momentum. $n_0$ is the condensate fraction, i.e., the fraction of
$^4$He atoms occupying the state with $p=0$. 

Evidently, the normalization of the density matrix given in
Eq. (\ref{norm}) implies, for the momentum distribution, the
normalization sum rule
\begin{equation}
 \frac{\nu}{(2\pi)^3\rho}\int d{\bf p}\ n(p)\ = \rho(0) = 1.
\label{normp}
\end{equation}  

The momentum distribution has been calculated in two different
manners. Having sampled the density matrix as function of $r$ one can
just take its Fourier transform according to
Eq. (\ref{momentum}). Alternatively, restricting to the RM method, one
can also directly accumulate
\begin{equation}
n({\bf p})=\frac{\rho}{\nu}\frac{1}{\cal N}\sum_{i=1}^{\cal N}
\frac{e^{i{\bf p}\cdot{\bf r}}\Psi(R'_i)}{\Psi(R_i)}.
\label{momentum1}
\end{equation}
At a first sight it seems  that the ${\bf r}$ integration is
missing in Eq. (\ref{momentum1}) above. However, a little reflection shows
that accumulating the estimator of Eq. (\ref{momentum1}) correctly
implements the integration (average) over ${\bf r}$, which appears in
the definition (\ref{momentum}) of the momentum distribution, since
${\bf r}$ is chosen at random for each configuration $R_i$, with a
uniform distribution in the simulation box. This second calculation of
$n({\bf p})$ is implemented for ${\bf p}$'s that are reciprocal
lattice vectors of the simulation cell. Eq. (\ref{momentum1}) is employed to accumulate variational and mixed
estimators, from which the extrapolated estimator is then obtained, as
explained above. From Eqs. (\ref{dmatrix1}) it follows that the
kinetic energy per particle $T$ can be related to the curvature of the
density matrix at the origin, according to
\begin{equation}
T= -\left[\frac{\hbar^2}{2M}\nabla^2 \rho(r)\right]_{r=0},
\label{kinetic1}
\end{equation}
which also implies the kinetic energy sum rule (see
Eq. (\ref{momentum}))
\begin{equation}
T=\frac{\hbar^2}{2M}\frac{\nu}{(2\pi)^3\rho}\int d{\bf p} p^2 n(p).
\label{kinetic}
\end{equation}

\section{Liquid $^4$He}

We have carried out DMC calculations of the one--body density matrix
and of the momentum distribution of $^4$He at four densities, using
EMC trial functions and 64 atoms in the simulation box. Selected runs
with up to 232 particles have been performed to check for finite size
effects.  In Fig.  \ref{fig1} we show our results for the one--body
density matrix $\rho(r)$ at the equilibrium density.  The kinetic
energy sum rule (\ref{kinetic}) is manifestly satisfied and the
saturation to a finite $n_0$ at large $r$ is evident, in spite of the
fact that with 64 particles only distances up to about $7\AA$ are
accessible. We also give in the inset a comparison between estimates
obtained with the FM and RM methods. The greater accuracy of the
latter method at large distances is apparent. In Fig. \ref{fig2} we
report, also at the equilibrium density, extrapolated estimates of the
momentum distribution $n(p)$, obtained using
Eq. (\ref{momentum1}). Due to the finite size of the system only wave
vectors larger than $\sim 0.4\AA^{-1}$ are accessible.  We note that a
shoulder is discernible in $pn(p)$ at $p\agt 2\AA$.
\par In order to extract the condensate fraction $n_0$ from our DMC
results, as well as to facilitate applications, we have fitted our DMC
extrapolated estimates of the density matrix and momentum
distribution, obtained with the RM method, to the following analytic
formula:
\begin{figure}
\null 
\psfig{figure=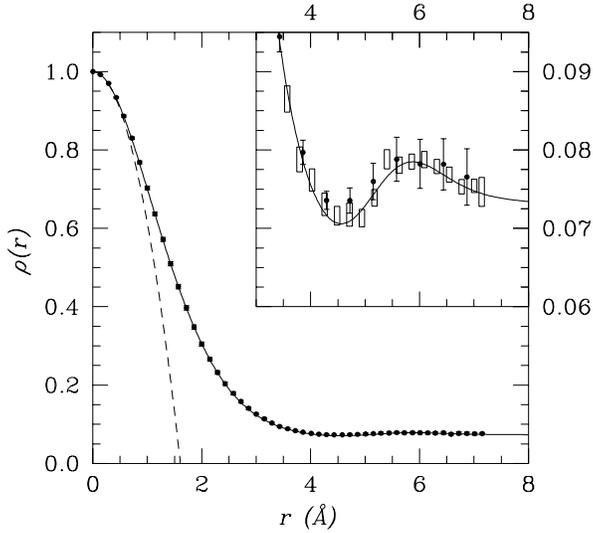,width=9cm}
\vspace{-1mm}
\caption[dum4] {One body--density matrix of $^4$He at the equilibrium
density $\rho(\AA^{-3})=0.02186$. The full curve is the fit of
Eq. (\ref{nk4fit}) to our DMC results (extrapolated estimates), with
the dashed curve showing the parabola $1-(MT/3\hbar^2)r^2$  that
satisfies the kinetic energy sum rule. The solid circles and
rectangles respectively give our DMC results obtained with the FM and
RM methods.}
\label{fig1} 
\end{figure}
\begin{eqnarray}
n(p)=&&(2\pi)^3\rho\delta({\bf p})n_0 \cr &&
+\left(n_0\frac{p_1}{p}+ n_1
\cos^2\left(\frac{p}{p_2}\right)\right)e^{-(p/p_3)^{\alpha}}
\cr &&+n_2e^{-(p-p_4)^2/p_5^2}.
\label{nk4fit}
\end{eqnarray}
The first 2 terms in Eq. (\ref{nk4fit}) account for the existence of
the condensate, while the third suitably models the shoulder in
$pn(p)$.  The fourth and last term accounts for the gross main
structure of the momentum distribution.  

We have simultaneously fitted $n(p)$ and $\rho(r)$, which must 
be obtained \hfil numerically by Fourier transforming 
the function of Eq. (\ref{nk4fit}), imposing as well the normalization
condition (\ref{normp}) and the kinetic energy sum rule
(\ref{kinetic}). Moreover, we set $p_1=Mc/2\hbar$ to satisfy the
long-wavelength behavior
\begin{table}
\caption[dumm1]{Parameters of the fit (\ref{nk4fit}) to the DMC
momentum distribution and density matrix of $^4$He at $T=0^o$K, at
various densities.  $\rho$ is in $\AA^{-3}$ and $p_1$--$p_5$ are in
$\AA^{-1}$.}
\begin{tabular}{c|llll}
   $\rho$ & 0.01964 & 0.02186 & 0.02401 & 0.02622\\ 
\hline 
$n_0$   &0.11163  &0.071673    &0.046227  &0.027079\\
$\alpha$ &  1.6941 & 1.7634   & 2.2342 & 1.9976\\
$n_1$ &  0.055274 & 0.038282   & 0.019157 & 0.021497\\
$n_2$ &  0.40285 & 0.39893  & 0.36695 & 0.38483\\
$p_1$    &  0.49957 & 0.73394   & 0.95934 & 1.1964 \\
$p_2$ &  0.38971 & 0.38525   & 0.40435 & 0.40588\\
$p_3$ &  1.4900 & 1.6744  & 2.2153 & 2.1832\\
$p_4$ &  0.29713 & 0.20538   & 0.28494 & 0.062765\\
$p_5$ &  0.85128 & 1.0016   & 1.0132 & 1.2549\\
\end{tabular}
\label{nkhe4}
\end{table} 

\begin{figure}
\null \vspace{-2mm}
\psfig{figure=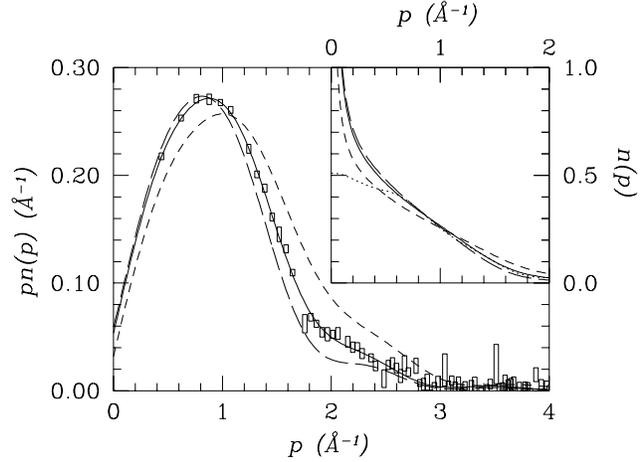,width=9cm}
\vspace{-8mm}
\caption[dum7]{\small Momentum distribution of $^4$He at
$T=0^o$K. Long dashes, full curve and short dashes are fits to the DMC
results at $\rho(\AA^{-3})= 0.01964, 0.02186, 0.02622$, whereas the
rectangles give the DMC extrapolated estimates at the equilibrium
density $\rho(\AA^{-3})=0.02186$. The inset shows the long--wavelength
behavior of $n(p)$, with the dotted curve reporting the results of
GFMC\cite{Whitlock87}.}
\label{fig2}
\end{figure}
\begin{equation}
 \lim_{p\rightarrow 0} p n(p)=\frac{n_0 M c}{2\hbar},
\label{longw}
\end{equation}
 induced by long--range correlations, as first discussed by
Gavoret and Nozieres\cite{Gavoret64}. Above $M$ is the mass of the
$^4$He atom and $c$ the sound velocity, which we estimate from the
DMC\cite{Moroni95b} equation of state (EOS). Thus we fit at each
density 6 independent parameters to more than 100 MC points, obtaining
a reduced $\chi^2$ between 0.98 and 1.20.  The
resulting fit parameters are recorded in Table \ref{nkhe4}.

\begin{figure}
\null
\psfig{figure=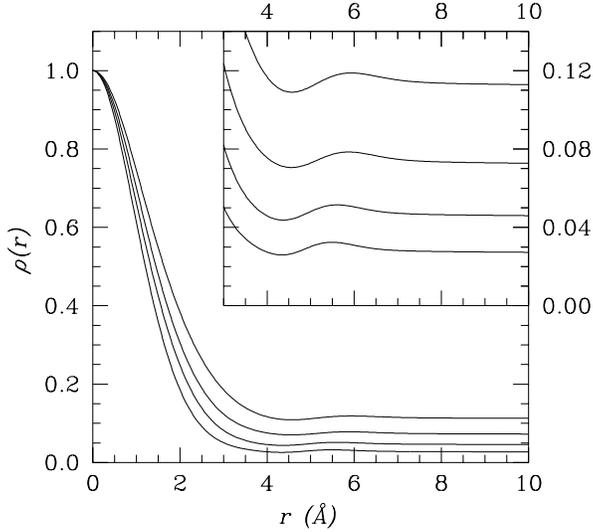,width=9cm}
 \caption[dum3] {One--body density matrix $\rho(r)$ of $^4$He. The
curves from the topmost to the lowest, give the analytical fit of
Eq. (\ref{nk4fit}) to our DMC results (extrapolated estimates),
respectively at $\rho(\AA^{-3})=0.01964$, $0.02186$, $0.02401$ and
$0.02622$. The inset shows $\rho(r)$ in the tail region with an
enlarged scale.}
\label{fig3} 
\end{figure}

In principle one could take $c$ too as unknown and get an independent
estimate of the sound velocity.  We have tried this alternative, at
the equilibrium density, obtaining an estimate of $c$ that is about
$10\%$ lower than experimental and DMC-EOS estimates, and has however
a very large uncertainty ( $\sim 50\%$), reflecting the absence of DMC
points for $p\alt 0.4 \AA^{-1}$), where the singular term in $n(p)$ is
important.  We may conclude that our data are compatible with the
presence of singular term in $n(p)$\cite{Gavoret64}, in spite of the lack of long--range terms in
the pseudopotentials that we have used\cite{Moroni95b}. In
Fig. \ref{fig3} we show the density matrix at the four densities that
we have studied, as given by the fit of Eq. (\ref{nk4fit}) with the
parameters of Table \ref{nkhe4}.

\begin{figure}
\vspace{-5mm}
\hspace{.5cm}\psfig{figure=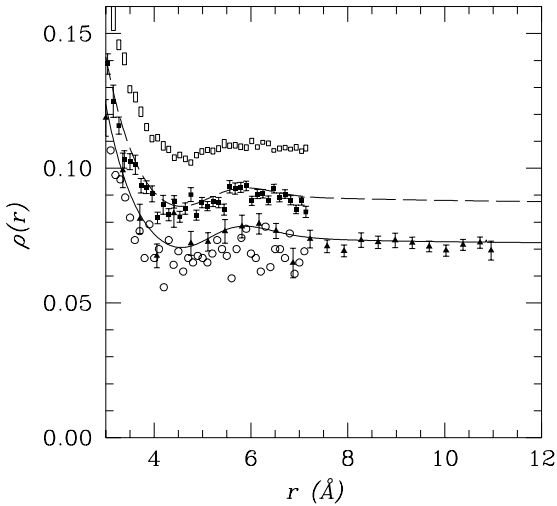,width=9cm}
\caption[dum5]{Dependence on the quality of the wave function and on
the method (VMC versus DMC) of the one body--density matrix of $^4$He,
at the equilibrium density $\rho(\AA^{-3})=0.02186$. The full curve is
the fit to our DMC results (extrapolated estimates) for 64 particles,
using our best trial function (OJOT), while the dashed curve gives the
fit to the VMC results obtained with this trial function. Empty
rectangles and solid squares give respectively VMC and DMC results
obtained from a simple OJ trial function, which  embodies only pair
pseudopotentials.  Finally, the triangles are the DMC results obtained
using the OJOT trial function and 232 particles and the circles report
the finding of PIMC at $T=1.18^o$K\cite{Ceperley86}.}
\label{fig4} 
\end{figure}

DMC momentum distribution and density matrix are compared with the fit
of Eq. (\ref{nk4fit}) in Figs. \ref{fig1}, \ref{fig2}, and \ref{fig4},
at the equilibrium density. The fit appears to be very good. A similar conclusion holds at the other densities that
we have studied.

The condensate fraction $n_0$ is mostly constrained by the large $r$
behavior of $\rho(r)$, which results into the term proportional to
$\delta({\bf p})$ in the momentum distribution. As we have already
mentioned, the singular behavior of $n(p)$ at small $p$, implied by
Eq. (\ref{longw}), is much less effective in 

\noindent determining $n_0$, because
of the absence of DMC estimates for $p\alt 0.4\AA^{-1}$.  From the
inset of Fig (\ref{fig1}) one might conclude that the range $r\alt 7
\AA$ accessible with 64 atoms is not big enough too unambiguously
assess the value of $n_0$. However, simulations with 232 atoms yield,
in the extra range $7\AA\alt r\alt 11$, DMC estimates that are in
perfect agreement with the fit to the $64$ particles results, as it is
clear from Fig. \ref{fig4}. We also illustrate in this figure the
dependence of the large $r$ limit of density matrix, i.e., $n_0$, on
the quality of the wave function. In particular it is apparent than
improving the MC description either  changing from VMC to DMC, for given
trial function, or changing to a better trial function in DMC, results
into a  decrease of $n_0$, in the case considered.

\begin{figure}
\hspace{.5cm}\psfig{figure=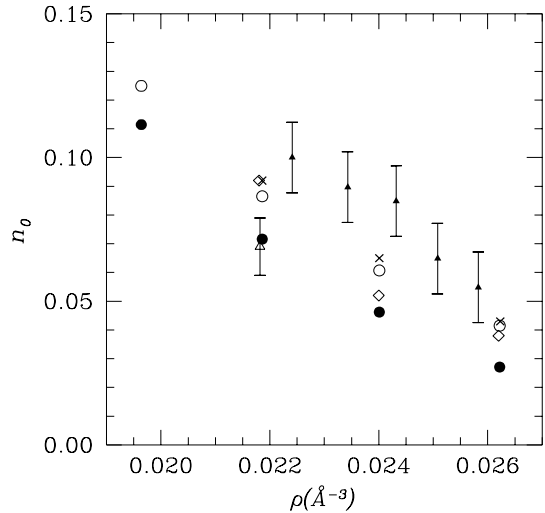,width=9cm}
\caption[dum6]{Condensate fraction of liquid $4$He, as a function of
the density. DMC (solid circles); EMC (open circles);
GFMC\cite{Whitlock87} (diamonds); HNC\cite{Manousakis85} (crosses);
experimental estimates at $T=0.75 ^o$K\cite{Snow92} (solid triangles); PIMC
at $T=1.18^o$K\cite{Ceperley86} (empty triangle).}  
\label{fig5} 
\end{figure}

In Fig. {\ref{fig5} and Table \ref{n0t} we compare our predictions for
$n_0$ with those from other theoretical treatments as well as with
experimental results at low temperature. Consistently with the
observation made above, our use of very accurate trial
functions\cite{Moroni95b} yields DMC predictions for the condensate
fraction which are lower than previously obtained by
GFMC\cite{Whitlock87} and HNC\cite{Manousakis85}. On the theoretical
side the only prediction that agrees with our own, though it has a
much larger statistical error, is the PIMC one at
$T=1.18^o$K\cite{Ceperley86}. We should remind the reader that in fact
PIMC has no trial function bias.  We find instead a sizeable
discrepancy with the experimental estimates of Snow {\it et
al}\,\cite{Snow92}, who determine $n_0$ by fitting a model $n(p)$ to the
measured Compton profile $J(Y)$---a procedure however which appears to
be model dependent. Different choices for   $n(p)$  produce
equivalently good fits of $J(Y)$\cite{Sokol95}, though embodying very
different condensate fractions, all the way from $n_0=0$ to
$n_0=10\%$. In fact , our DMC momentum distribution yields prediction
for the Compton profile which agrees well with the experiments, as we
show below.

\begin{figure}
\hspace{.5cm}\psfig{figure=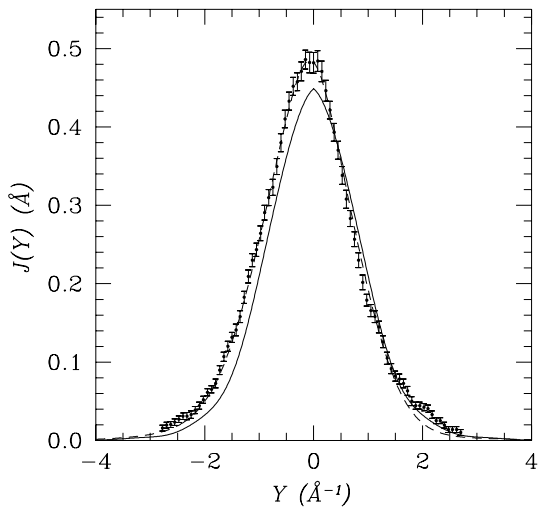,width=9cm}
\caption[dum8]{Compton profile $J(Y)$ of $^4$He at the
equilibrium density $\rho(\AA^{-3})=0.02186$ compared with
experimental (solid circles) data at $Q=23\AA^{-1}$ and
$T=0.35^oK$\cite{sss90}. The dashed (solid) curve is obtained from our
fit to the DMC $n(p)$ allowing (not allowing) for both the
experimental broadening and the final state interactions of
Ref. \onlinecite{sss90}. In calculating the dashed curve a shift of
$-0.1\AA$ in $Y$ has been also used as in \onlinecite{sss90}.  }
\label{fig6}
\end{figure}

The inelastic neutron scattering cross--section at high momentum
transfer $\hbar Q$ can be approximated by its Impulse Approximation (IA)
expression, which is proportional to the Compton profile\cite{Sokol95}
\begin{equation}
J(Y)=\frac{1}{4\pi^2\rho}\int_{|Y|}^{\infty} dp\ pn(p).
\label{compton}
\end{equation}
The scattering, in IA, does not depend on the energy $\omega$ and the
momentum transfer $\hbar Q$ separately, but only through the scaling
variable $Y$, given by
\begin{equation}
Y=(M/\hbar Q)(\omega-\omega_r),
\label{scaling}
\end{equation}
where $\omega_r=\hbar^2 Q^2/2M$ is the recoil energy of the scattering
atom.  The dynamical response function $S(Q,\omega)$, in IA, is given
by $J(Y)$ times the factor $M/(\hbar Q)$. Final state effects (FSE) of
the medium on the scattered atom as well 
\begin{table}
\caption[dummy]{Condensate fraction $n_0$ in $^4$He. DMC,
GFMC\cite{Whitlock87}, and HNC\cite{Manousakis85} predictions are at $T=0$. The
PIMC\cite{Ceperley86} result is at $T=1.18^o$K and  the density is in
$\AA^{-3}$.  The figure in parenthesis is the uncertainty on the last
figure, whenever available.}
\begin{tabular}{c|lllll}
$\rho$ & 0.01964 & 0.02186 & 0.02401 & 0.02622\\ 
\hline 
DMC &  0.112(1) & 0.0717(5) & 0.0462(6)& 0.02.71(6) \\ 
PIMC&                & 0.069(10)       &&\cr   
GFMC &               & 0.092(1)   & 0.052(1)     & 0.037(2)\\
HNC &               & 0.092       & 0.065        & 0.043 \\
\end{tabular}
\label{n0t}
\end{table} 
\noindent
\noindent  as experimental
resolution (ER) broaden up the Compton profile, particularly its delta
peak at $p=0$, which is due to the Bose condensation.  In
Fig. \ref{fig8} Compton profiles, calculated with our DMC momentum
distributions are compared with observed scattering data\cite{sss90}
at $T=0.35^o$K, converted to $J(Y)$. Once ER and FSE are taken into
account\cite{sss90}, good agreement with the experiment is obtained .

\section{Liquid $^3$He}

For normal $^3$He, DMC simulations using EMC trial functions with
backflow and the fixed--node approximation have been performed at
five densities, with 54 atoms in the simulation box. We have
investigated the dependence of the momentum distribution on the size
of the system and on the range of the backflow, respectively with runs
for 114 atoms and with runs using trial functions embodying long-range
backflow.

\begin{figure}
\psfig{figure=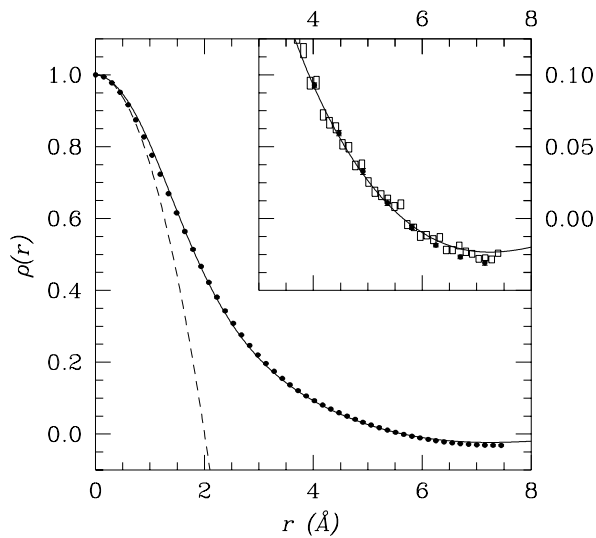,width=9cm}
\caption[dum10]{One body--density matrix of $^3$He at the equilibrium
density $\rho(\AA^{-3})=0.01635$. The full curve is the fit of
Eq. (\ref{nk3fit}) to our DMC results (extrapolated estimates), with
the dashed curve showing the parabola $1-(MT/3\hbar^2)r^2$ that
satisfies the kinetic energy sum rule. The solid circles and
rectangles respectively give our DMC results obtained with the FM and
RM methods.}
\label{fig7} \end{figure}

In Fig. \ref{fig7} we give our DMC estimates for the
density matrix at the equilibrium density. It is clear that the
kinetic energy sum rule (\ref{kinetic}) is satisfied. The size of the system
allows for the determination of $\rho(r)$ through its first zero and
up to the first minimum.  A comparison between results obtained with
FM and RM methods is  also given, in the inset. The smaller error on
the FM results, compared with that on the RM estimates, is due to the
much longer runs used to accumulate the FM $\rho(r)$ in this case.

\begin{figure}
\null
\vspace{3mm}
\psfig{figure=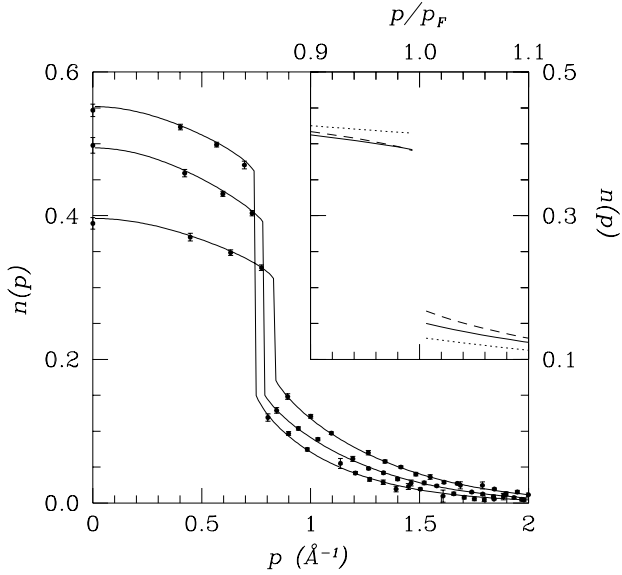,width=9cm}
\vspace{-1mm}
\caption[dum11]{ Momentum distribution of $^3$He at $T=0^o$K and
$\rho(\AA^{-3})=0.01413$, $0.01635$ and $0.01946$: dots are DMC
(fixed--node) extrapolated estimates; full curves are fits to the DMC
results.  Larger densities correspond to lower curves at $p=0$.  The
inset shows the behavior around the Fermi momentum at the equilibrium
density $\rho(\AA^{-3})=0.01635\AA^{3}$, using PW ($\cdots$), ~SBF (---) , and
LBF (- - -) trial functions.}  
\label{fig8} 
\end{figure}

In Fig. \ref{fig8} we report, at three different densities, the
momentum distribution $n(p)$ obtained using
Eq. (\ref{momentum1}). Clearly, the discontinuity $Z$ at the Fermi
wave vector $p_F$ is substantially reduced, with respect to its value
$Z=1$ in the noninteracting system, and moreover it systematically
shrinks, as the system gets denser and the effects of the interaction
become more important.  Z is also  slightly reduced
\begin{figure}
\vspace{-2mm}
\hspace{.5cm}\psfig{figure=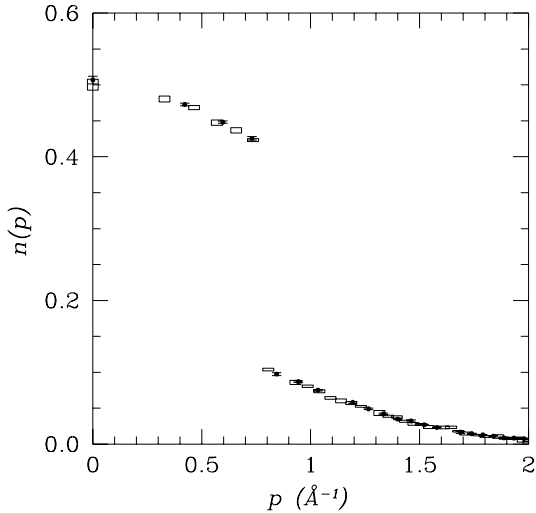,width=9cm}
\vspace{-3mm}
\caption[dum12]{Momentum distribution of $^3$He at equilibrium density
$\rho(\AA^{-3})=0.01635$ with the VMC method, using OJ trial
functions, with 54 (solid circles) and 114 (rectangles) particles.}
\label{fig9} 
\end{figure}
\noindent 
 when the nodes of
the trial function are improved from PW to SRB and then to LBF.  Size
effects on the momentum distribution appear to be negligible, as it is
clear from the comparison between variational results for 54 and 114
particles given in Fig. \ref{fig9}, at the equilibrium density.

In a normal Fermi liquid such as $^3$He the momentum distribution, in
addition to the discontinuity, has infinite
slopes\cite{Fantoni83,Sartor80} at $p_F$. To leading order in $p-p_F$
\begin{eqnarray}
n(p\rightarrow p_F^{\pm}) & \simeq & 
n(p_F^{\pm})+A\frac{p-p_F}{p_F}\ln\left|\frac{p-p_F}{p_F}\right|,
\label{singular} 
\end{eqnarray}
with the  coefficient $A$  related to imaginary part of the
self--energy $\Sigma({\bf p},E)$. 

To  extract   $Z$ from the calculated momentum distribution and
to check that our results are consistent with the presence of the singular
term of Eq. (\ref{singular}) we have fitted our data for $\rho(r)$ and
$n(p)$ to the real space form 
\begin{eqnarray}
\rho(r)=&&Z\rho_0(x) + a_1\frac{\rho_0(x)}{x} +
a_2\frac{x\rho_0(x)-sin(x)}{x^3} \cr &&-a_1\frac{e^{-b_1x}}{x}
+(a_3+a_4 x+a_5x^2+a_6x^3)e^{-b_2x} 
\cr
 &&
\label{nk3fit}
\end{eqnarray}
with $x=p_Fr$, and 
\begin{equation}
\rho_0(r)=\frac{3}{x^3}(\sin x-x\cos x),
\label{rhou}
\end{equation}
the density matrix of the ideal Fermi Gas. The first two terms in
Eq. (\ref{nk3fit}) account respectively for the discontinuity and the
infinite slope of $n(p)$ at $p_F$. The third term allows for a finite
discontinuities in the first and second derivatives of $n(p)$ at
$p_F$. Finally, the fourth term is needed to eliminate the divergence
that the term $\rho_0(x)/x$ produces at the origin. We impose the
normalization condition (\ref{normp}) and the kinetic energy sum rule
(\ref{kinetic}), as well as the vanishing of the first and third
derivative of $\rho(r)$ at $r=0$. Thus we fit at each density 5
independent parameters to more than 100 MC points, with a reduced
$\chi^2$ between 0.87 and 1.15. The resulting fit parameters are recorded in
Table \ref{nkhe3}.

We compare fit and DMC estimates for $\rho(r)$ and for $n(p)$,
respectively in Fig. \ref{fig7} and Fig. \ref{fig8}. Evidently,
Eq. (\ref{nk3fit}) is fully consistent with our DMC results.  In
Fig. \ref{fig10} we show the one--body density matrix at three of the
densities that we have studied, as given by the fit of
Eq. (\ref{nk3fit}).  As for the non--interacting case, $\rho(r)$
becomes steeper with increasing the density and its first zero moves
toward the origin.

A comparison of our results for the momentum distribution of $^3$He
with those from some other calculations is given in Fig. \ref{fig11} and in
Table \ref{zt}. HNC results\cite{Fabrocini92} are in 
close agreement with our variational $n(p)$, for $p\geq p_F$ as well
as with the estimate of $Z$, while small differences are present at
small momenta, which are however of little relevance in the density of
states $\propto n(p)p^2$.  The evident discrepancies between our
results and those of GFMC\cite{Sokol85}
\begin{figure}
\null
\vspace{-3mm}
\hspace{.5cm}\psfig{figure=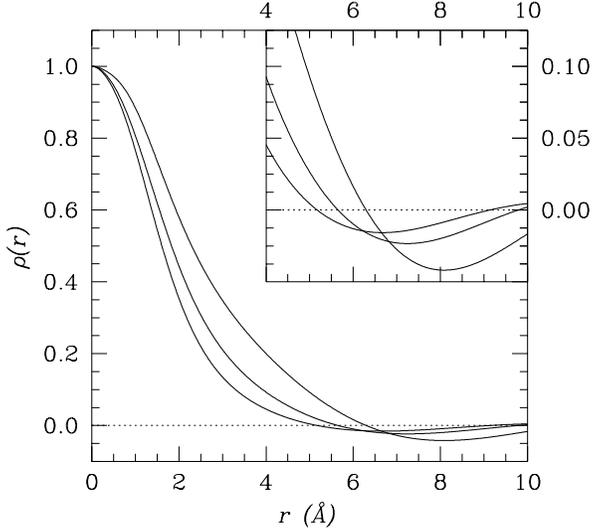,width=9cm}
  \caption[dum9]{One--body density matrices $\rho$ of $^3$He. The
curves from the less to the most steep, give the analytical fit of
Eq. (\ref{nk3fit}) to our fixed-node DMC results (extrapolated
estimates), respectively at $\rho(\AA^{-3})=0.01198$, $0.01635$ and
$0.01946$.The inset shows $\rho(r)$ in the tail region with an
enlarged scale. }
\label{fig10} 
\end{figure}
\noindent
\noindent  around the Fermi surface are
probably due to the poorer trial function used in GFMC\cite{Sokol85},
as well as to statistical errors.

As we have already mentioned, the functional form of
Eq. (\ref{nk3fit}) implies for $n(p)$ a singular term at $p_F$ of the
form (\ref{singular}), with $A=a_1/\pi$.  For instance, at the
equilibrium density $\rho(\AA^{-3})=0.01635$ we find $A=0.06(2)$,
being however unable at present to assess the size dependence of such
estimate. An 
independent estimate of $A$ is given by a perturbation
calculation\cite{Sartor80} for a dilute, hard--sphere, Fermi gas.  To
order $(p_FR)^2$,
\begin{equation}
A_0=2\frac{\nu-1}{\pi^2}(p_FR)^2,
\end{equation}
with $R$ the radius of
the Fermi particle. Evidently $^3$He at equilibrium is no at all
dilute. Nevertheless, taking $R=\sigma/2\simeq 1.3\AA$, one gets
$A_0\simeq 0.20$, which is of the
same order of magnitude as our DMC
estimate. In CBF 
\vspace{-2mm}
\begin{table}
\caption[dumm2] {Parameters of the fit (\ref{nk3fit}) to the
fixed--node DMC momentum distribution and density matrix of $^3$He at
$T=0^o$K, at various densities.  $\rho$ is in $\AA^{-3}$.}
\begin{tabular}{c|lllll}
 $\rho$  & 0.01198  &0.01413  &0.01635  &0.01797  &0.01946\\
\hline 
$Z$   &0.45977  &0.30598  &0.23616  &0.14328  &0.13566 \\
$a_1$ &0.14248  &0.28612  &0.19376  &0.41699  &0.24613 \\
$a_2$ &\hspace{-1.3mm}-0.11973 &\hspace{-1.3mm}-0.21912 
&\hspace{-1.3mm}-0.12381 &\hspace{-1.3mm}-0.28125 &\hspace{-1.3mm}-0.23982 \\
$a_3$ &\hspace{-1.3mm}-0.73200 &\hspace{-1.3mm}-0.80732 
&\hspace{-1.3mm}-0.64305 &\hspace{-1.3mm}-2.1780 &\hspace{-1.3mm}-0.96272 \\
$a_4$ &3.2095   &1.9033   &3.2723   &4.1029  &3.9981 \\
$a_5$ &\hspace{-1.3mm}-1.5905  &0.64801  &\hspace{-1.3mm}-0.098765 
&\hspace{-1.3mm}-2.4600 &\hspace{-1.3mm}-0.64725 \\
$a_6$ & 7.6229  &1.5582   &4.6639    &6.8026  &6.2113 \\
$b_1$ & 9.0415  &5.3494   &7.3460   &7.3676   &7.5530 \\
$b_2$ & 3.5907  &2.7487   &3.0716   &3.3316   &3.1652 \\
\end{tabular}
\label{nkhe3}
\end{table} 
\begin{figure}
\null
\vspace{-3mm}
\hspace{.5cm}\psfig{figure=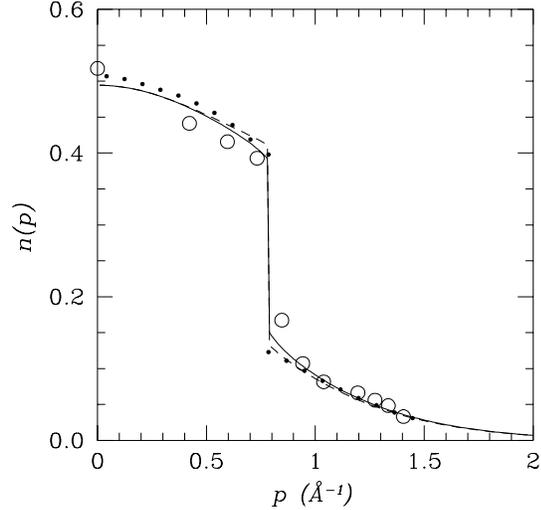,width=9cm}
 \caption[dum13]{Momentum distribution of $^3$He at equilibrium
density $\rho(\AA^{-3})=0.01635$: comparison of the present DMC (---)
and VMC (- - -) fits with the predictions of
GFMC\cite{Whitlock87}(circles) and HNC\cite{Fabrocini92} (solid
circles).}
\label{fig11} \end{figure}
\noindent theory one obtains\cite{Fantoni84}, instead,  
\begin{equation}
A_{CBF}=\frac{2W_0}{\pi}p_F\frac{d\ e^v(p_F)}{d\ p},
\label{cbfrelat}
\end{equation}
where $e^v(p)$ is the variational energy and $W_0$ is an inverse
energy parameter characterizing the imaginary part $W(p,E)$ of the
self--energy $\Sigma(p,E)$, close to the Fermi energy,
\begin{equation}
W(p,E)\simeq W_0(E-e_F)^2, \ \ \    E\rightarrow e_F. 
\end{equation}
If one takes $W_0\simeq 2.5 ^oK^{-1}$ and
$de^v(p_F/dp=\hbar^2p_F/M_v$, $M_v/M=0.76$, from earlier
work\cite{Fantoni82} where a model form of $W(p,E)$ was fitted to the
measured specific heat in $^3$He, $A_{CBF}\simeq 21$ is obtained,
which is 2 order of magnitude larger than  both the DMC and
the  perturbative  estimate.  

The effective mass $M^*$ is related to the dispersion at $p_F$ of the
quasi--particle energy\cite{agd63}
\begin{equation}
e(p)=\frac{\hbar^2p^2}{2M} +\Re\Sigma(p,e(p)), 
\label{singlepar}
\end{equation}
according to $\hbar^2 p_F/M^*=de(p_F)/dp$.
Thus\cite{Jeukenne86,Fantoni84,Fabrocini92} $M^*/M = M_EM_K$ with the
K--mass
\begin{equation}
 M_K^{-1} = 1+\frac{M}{\hbar^2 p_F}\frac{\partial}{\partial p}
\Re\Sigma(p,E)\mid_{E=e_F,p=p_F}.  
\label{kmass}
\end{equation}
and the E--mass
\begin{equation}
M_E= 1-\frac{\partial}{\partial E} \Re \Sigma({\bf
 p},E)\mid_{E=e_F,p=p_F} =Z^{-1}. \label{emass}
\end{equation}
Thus, the E--mass is nothing but $Z^{-1}$, i.e., the strength of the
quasi--particle pole at $p_F$\cite{Migdal57,Luttinger60}. We lack,
however, an estimate of the K--mass, to predict $M^*$. Hence, we are
planning to perform variational and transient estimate calculations of
$M^*$, along the lines of an equivalent  calculation for the 2--dimensional
electron gas\cite{kcm94,kcm96}.

\begin{figure}
\vspace{-1mm}
\hspace{.5cm}\psfig{figure=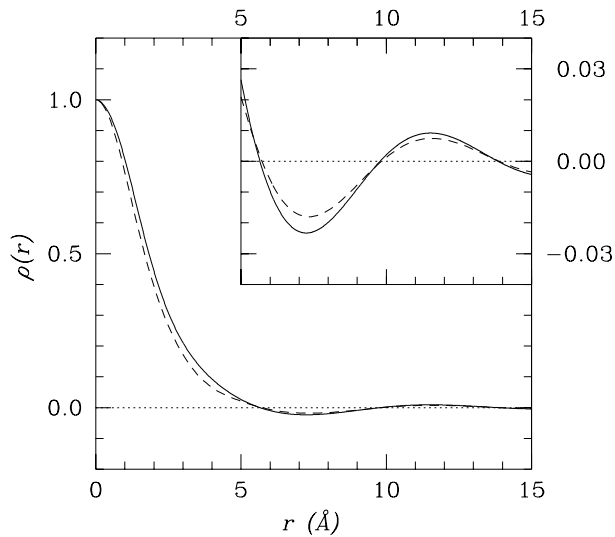,width=9cm}
\vspace{-4mm}
 \caption[dum14]{
The one-body density matrix $\rho(r)$ of $^3$He at $T=0^o$K and
$\rho_e=0.01635 \AA^3$ (full curve). The dashed curve gives the
product $\rho_0(r)\rho_B(r)$, with $\rho_0(r)$ and $\rho_B(r)$
respectively the density matrix for the ideal (uncorrelated) Fermi gas
and the density matrix of a  Bose $^3$He.}
\label{fig12} \end{figure}

At the Fermi wave vector $p_F$, the momentum distribution $n(p)$ has a
discontinuity and, according to perturbation theory and to CBF, at
least the additional singular behavior of Eq. (\ref{singular}). It is
known\cite{l59} that singularities dominate the large distance
behavior of the Fourier transform of a generalized function, such as
$n(p)$.  This imply in particular that as $r\rightarrow \infty$
$\rho(r)\simeq Z \rho_0(r)$, to leading order. On the other hand,
using the parameters given in Table \ref{nkhe3} one can show that the
first zero of the DMC density matrix at the equilibrium density (see,
also, Fig. \ref{fig7}) is at $p_Fr=5.62$, which is not very different
from $p_Fr=5.72$---the location of the first zero of $\rho_0(r)$. We
therefore consider the approximate decoupling
\begin{equation}
\rho(r)= \rho_0(r)\rho_B(r), 
\label{factor}
\end{equation}
with the function $\rho_B(r)$ that must satisfy $\rho_B(0)=1$,
$\lim_{r\rightarrow \infty} \rho_B(r) = Z$, and we further choose it to
be everywhere non--negative. It is tempting to take $\rho_B(r)$ as the
density matrix of a suitable Bose system, with a condensate fraction
$Z$.

We have thus simulated a system of $^3$He with Bose statistics, at the
equilibrium density of real $^3$He. We find a condensate fraction of
$0.208(5)$, which within error bars agrees with our best estimate of
$Z=0.21(2)$.  In Fig. \ref{fig12} we compare our fit to the density
matrix of $^3$He with the prediction of the approximate formula
(\ref{factor}), using for $\rho_B(r)$ the fit to the simulated density
matrix of a Bose $^3$He.  It is apparent that the decoupling of
Eq. (\ref{factor}) approximately holds. In fact Eq. (\ref{factor})
overestimates the kinetic energy by $14\%$, while underestimating the
envelope of the tail of $\rho(r)$ by about a $20\%$. Thus statistics
and correlations in $^3$He decouple, within a reasonable accuracy,
into those of an ideal (uncorrelated) Fermi gas and a Bose $^3$He, as
far as $\rho(r)$ is concerned.

\section{Conclusions}

We have presented in this paper results of DMC calculations of the
one--body density matrix and the momentum distribution of liquid
$^4$He and liquid $^3$He. These DMC simulations are based on accurate
trial wave functions with fully optimized pair and triplet
pseudopotentials. For $^3$He the Fermion sign problem has been
avoided, by resorting to the fixed--node approximation. In particular,
we have used backflow nodes, which are more accurate than the simpler
plane--wave nodes. We have recorded our data in a form suitable for
future use, in terms of analytical fits.

Our prediction for the condensate fraction of $^4$He is lower than in
previous microscopic calculations and we have argued that these seems
to be consistently related to the improved description of $^4$He
afforded by the calculations presented here. We have also found that our
results are statistically consistent with the presence of a singular
term in $n(p)$ as  predicted earlier by Gavoret and Nozieres, in
spite of the absence of explicit long--range correlations in our trial
functions.

The discontinuity $Z$ of $n(p)$ at $p_F$ in $^3$He is also predicted
from our calculations to be sensibly lower than in previous
variational calculation and in substantial agreement with GFMC
estimates. We have investigate the presence in the $n(p)$ of
logarithmic singularities, as predicted by approximate treatments.  We
find that our results are compatible with the presence of such
terms. However the strength of such term is in order of magnitude
agreement with perturbation theory, whereas a CBF treatment with
empirical parameters for the imaginary 
part of the optical potential
imply a strength which differs from our prediction by 2 orders of
magnitude.

We have have also demonstrated that an approximate decoupling
\hfil $\rho(r)\simeq \rho_0(r)\rho_B(r)$ \hfil ~holds, \hfil ~with \hfil
$\rho_0(r)$ \hfil ~and 

\begin{table}
\caption[dumm3] {Discontinuity of the momentum distribution at $p_F$,
$Z$, as function of the density, from various calculations. SBF and LBF
denote the DMC estimates, obtained using the fit of Eq. (\ref{nk3fit})
and results for EMC wave functions with short and long--range backflow,
respectively. VMC gives the variational estimate for the SBF wave function.}
\begin{tabular}{c|lllll} 
$\rho$  & 0.01198  &0.01413  &0.01635  &0.01797  &0.01946\\
\hline
HNC\tablenotemark[1] &          &0.348    &0.275    & 0.244   & 0.221\\
VMC     &          &         &0.272(2) &         &\\
GFMC\tablenotemark[2] &         &         &$\alt$ 0.2 &         & \\
SBF     & 0.46(2)  &0.31(2)  &0.24(1)  & 0.14(1) &0.14(1)\\
LBF     &          &         &0.21(2)  &         & \\
\end{tabular}  
\tablenotetext[1]{Ref.\ \onlinecite{Fabrocini92}}
\tablenotetext[2]{Ref.\ \onlinecite{Whitlock87} }
\label{zt} 
\end{table} 
\noindent $\rho_B(r)$ respectively the density matrix for the ideal
Fermi  gas and the density matrix of a Bose $^3$He.  Thus
statistics and correlations effects seem to decouple in $^3$He, as far
as the density matrix is concerned. The Bose $^3$He has in fact a
condensate fraction which agree within error bars with the
discontinuity $Z$ found in the Fermi $^3$He.

We believe that the estimates given in this paper provide the most
accurate information of this kind on He available to date. This is of
particular importance for the condensate fraction in $^4$He, as its
extraction from deep inelastic neutron scattering seems still not
feasible. Using our accurate EMC trial functions we have also studied
partially polarized $^3$He. We shall report on this study
elsewhere\cite{Moroni95c}.

\acknowledgments Most of the work presented in this paper was done
when SM was enjoying a postdoctoral fellowship at the Laboratorio FORUM of
the Istituto Nazionale di Fisica della Materia, Pisa. Access to the
computing facilities of CNUCE are also gratefully acknowledged.


\begin{thebibliography}{1}
 
\bibitem{momdis89} {\it Momentum Distributions}, R.N.Silver and
P.E.Sokol, eds (Plenum, N.Y., 1989).

\bibitem{agd63} See, e.g., A.L. Fetter and J.D. Walecka, 
{\it Quantum Theory of Many-Particle Systems} (McGraw-Hill, New York, 1971).

\bibitem{Ceperley79} D.M.Ceperley and M.H.Kalos, in {\it Monte Carlo
Methods in Statistical Physics}, K.Binder, ed. (Springer, N.Y., 1979).
 
\bibitem{Fantoni78} S.Fantoni, {\it Nuovo Cimento} {\bf 44A}, 191
(1978).

\bibitem{Ristig81} M.L.Ristig, in {\it From Particle to Nuclei},
Course LXXIX Int. School of Phys.{\it Enrico Fermi}, A.Molinari ed.,
(North--Holland, Amsterdam, 1981)

\bibitem{Fantoni84} S.Fantoni and V.R.Pandharipande, {\it Nucl. Phys.}
{\bf A427}, 473 (1984).

\bibitem{Manousakis85} E.Manousakis, V.R.Pandharipande and Q.N.Usmani,
{\it Phys. Rev.} {\bf B31}, 7022 (1985).

\bibitem{Fabrocini82} A. Fabrocini, S. Rosati, {\it Nuovo Cimento}
{\bf D1}, 567, 615 (1982).
  
\bibitem{Flynn86} M.F.Flynn, {\it Phys. Rev.} {\bf B33}, 91 (1986).

\bibitem{Fabrocini92} A.Fabrocini, V.R.Pandharipande and Q.N.Usmani,
{\it Nuovo Cimento} {\bf 14D}, 469 (1992).

\bibitem{Feenberg69} E. Feenberg, {\it Theory of Quantum Liquids}
(Academic, New York, 1969).


\bibitem{Usmani82} Q. N. Usmani, S. Fantoni, V. R. Pandharipande, {\it
Phys.  Rev.} {\bf B26}, 6123 (1982), and references therein.
 

\bibitem{Smith76} R. A. Smith, {\it Phys. Lett.} {\bf 63b}, 369
(1976); {\bf 85B}, 183 (1976); R. A. Smith, A. Kallio, M. Puoskari,
P. Toropainen, {\it Nucl.  Phys.} {\bf 328A}, 186 (1979).

  
\bibitem{Krotscheck86} E. Krotscheck, {\it Phys. Rev.} {\bf B33}, 3158
(1986).


\bibitem{Wang93} X.Q.G.Wang, S.Fantoni, E.Tosatti and L.Yu, {\it Phys
Rev} {\bf B49}, 10027, 1994.


\bibitem{Moroni95a} S.Moroni, S.Fantoni, G.Senatore, {\it
Europhys. Lett.}, {\bf 30}, 93 (1995).


\bibitem{Moroni95b} S.Moroni, S.Fantoni, G.Senatore, {\it Phys. Rev.},
{\bf B52}, 13547 (1995).

\bibitem{Whitlock87} P.A.Whitlock and R.Panoff, {\it Ca. J. of Phys.}
{\bf 65}, 1409 (1987).


\bibitem{Ceperley86} D.M.Ceperley and E.L.Pollock, {\it
Phys. Rev. Lett.} {\bf 56}, 351 (1986); {\it Can J. of Phys.} {\bf
65}, 1416 (1986); E.L.Pollock and D.M.Ceperley, {\it Phys. Rev.} {\bf
B36}, 8343 (1986).

\bibitem{Masserini92} G.L.Masserini, L.Reatto, S.A.Vitiello, {\it
Phys. Rev.  Lett.} {\bf 69}, 2098 (1992).


\bibitem{Reynolds82} P.J.Reynolds, D.M.Ceperley, B.J.Alder,
W.A.Lester, {\it J.  Chem. Phys.} {\bf 77}, 5593 (1982).


\bibitem{Ceperley91} D. M. Ceperley, in {\it Recent Progress in
Many--Body Theories}, ed. J. Zabolitzky (Springer 1981);
D. M. Ceperley, J.  Stat. Phys. {\bf 63}, 1237 (1991).


\bibitem{lkc74} K.S. Liu, M.H. Kalos, and G.V.Chester, Phys. Rev. {\bf
A10}, 303, (1974).
 
\bibitem{cb95} J. Casulleras and J. Boronat, Phys. Rev. {\bf B 52},
3654, (1995).

 
\bibitem{Vitiello92} S. A. Vitiello, K. E. Schmidt, {\it Phys. Rev.}
{\bf B46}, 5442 (1992).


\bibitem{Aziz79} R. A. Aziz, V. P. S. Nain, J. S. Carley,
W. L. Taylor, G. T.  Conville, {\it J. Chem. Phys.} {\bf 70}, 4330
(1979).



\bibitem{Schmidt81} K. E. Schmidt, M. A. Lee, M. W. Kalos,
G. V. Chester, {\it Phys. Rev. Lett.} {\bf 47}, 807 (1981).


\bibitem{Manousakis83} E. Manousakis, S. Fantoni, V. R. Pandharipande,
Q. N.  Usmani, {\it Phys. Rev.} {\bf B28}, 3770 (1983).


\bibitem{Kwon93} Y. Kwon, D. M. Ceperley, R. M. Martin, {\it
Phys. Rev.} {\bf B48}, 12037 (1993).


\bibitem{Panoff89} R.M.Panoff, J.Carlson, {\it Phys. Rev. Lett.} {\bf
62}, 1130 (1989).

\bibitem{Gavoret64} J.Gavoret and P.Nozi\'{e}res, {\it Ann. Phys.}
(NY) {\bf 28},349 (1964);
 
\bibitem{Snow92} W.N.Snow,Y.Wang and P.E.Sokol, {\it Europhys. Lett.}
{\bf 19}, 403 (1992)


\bibitem{Sokol95} P.E.Sokol in {\it Bose--Einstein condensation},
A.Griffin, D.W.Snoke and S. Stringari, eds. (Cambridge University
Press, 1995).


\bibitem{sss90} T.R.Sosnick, W.M.Snow, P.E.Sokol and R.N.Silver,
Europhys. Lett. {\bf 9}, 707 (1989).


\bibitem{Fantoni83} S.Fantoni, B.L.Friman and V.R.Pandharipande, {\it
Nucl.  Phys.} {\bf A399}, 51 (1983).


\bibitem{Sartor80} R.Sartor and C.Mahaux, {\it Phys. Rev.} {\bf C21},
1546 (1980).


\bibitem{Sokol85} P.E.Sokol, K.Sk\"{o}ld, D.L.Price, R.Kleb, {\it
Phys. Rev.  Lett.} {\bf 54}, 90 (1985).


\bibitem{Fantoni82} S.Fantoni, V.R.Pandharipande and K.E.Schmidt, {\it
Phys.  Rev. Lett.} {\bf 48}, 878 (1982).



\bibitem{Jeukenne86} J.P.Jeukenne, A.Lejeunne and C.Mahaux, {\it
Phys.Rep.} {\bf 25}, 83 (1986).

 
\bibitem{Migdal57} A.B.Migdal, {\it JETP (Sov. Phys)} {\bf 5}, 333
(1957).


\bibitem{Luttinger60} J.M.Luttinger, {\it Phys. Rev.} {\bf 119}, 1153
(1960).


\bibitem{kcm94}
Y. Kwon, D.M. Ceperley, and  R.M. Martin, Phys. Rev. {\bf B50}, 1684
(1994).

\bibitem{kcm96}
Y. Kwon, D.M. Ceperley, and  R.M. Martin, Phys. Rev. {\bf B53} 7376 (1996). 


\bibitem{l59}
M.J. Lighthill, {\it Introduction to Fourier Analysis and Generalized
Functions} (Cambridge University Press, Cambridge 1959).

\bibitem{Moroni95c} S. Moroni, S.Fantoni, G.Senatore, in preparation.

 \end{thebibliography}
\end{document}